\documentclass{ws-ijmpc}

\usepackage{subfigure, amssymb, amsmath}

\newcommand{\marketshare}[1][\alpha]{\langle c_{#1}\rangle}
\newcommand{\itemquality}[1][\alpha]{\langle l_{#1}\rangle}

\begin{document}
\markboth{Ama\c{c} Herda\u{g}delen, Haluk Bingol}
{A Cultural Market Model}


\title{A Cultural Market Model}
\author{Amac Herdagdelen}
\address{Computer Engineering Department, Bogazici University\\
Istanbul, Turkey\\
amac@herdagdelen.com}
\author{Haluk Bingol}
\address{Computer Engineering Department, Bogazici University\\
Istanbul, Turkey\\
bingol@boun.edu.tr}

\maketitle

\begin{history}
\received{19 September 2007}
\accepted{11 October 2007}
\end{history}

\begin{abstract}
Social interactions and personal tastes shape our consumption behaviors of cultural products. In this study, we present a computational model of a cultural market and we aim to analyze the behavior of the consumer population as an emergent phenomena. Our results suggest that the final market shares of the cultural products dramatically depend on the consumer heterogeneity and social interaction pressure. Furthermore, the relation between the resulting market shares and the social interaction is robust with respect to a wide range of variation in the parameter values and the type of topology.


\end{abstract}

\ccode{PACS Nos.: 87.23.Ge, 89.65.Gh}


\section{Introduction and Relevant Studies}

Social interaction is an inevitable aspect of our lives and a very strong ingredient of our decision processes. Most of our decisions depend, at least partially, on what other people think and how they behave. The extent of the society's influence on our behaviors may range from daily decisions such as what to wear at work to political decisions such as which party to vote for in the elections. The importance of the social interaction has been reflected in the social sciences for many decades and a growing body of research continues on the intersection of various disciplines including but not limited to sociology, cognitive sciences, physics and economics.~\cite{Farrell1985,Gordon2005,Granovetter1978,Markose2006,Schelling1973,Schelling1978} Statistical physics has a long history of dealing with interacting particles and emergent phenomena. Some of the techniques employed by statistical physics are also applied to human populations successfully and offer us new ways to explore the dynamics in social systems.~\cite{Durlauf1999,Galam1997,Holyst2000,Michard2005,Phan2006,Sznajd2000}

The interaction between the macro dynamics and individual decision processes is an active area of research in computational sociology and statistical physics (i.e. sociophysics). Problems in which the agents in a community are faced with a binary decision such as to vote for or against a legislation or to buy a particular product or not are extensively studied by the statistical physicists.~\cite{Gordon2005,Granovetter1978,Galam1997,Michard2005,Sznajd2000} The \textit{Random Field Ising Model} (RFIM) is a commonly investigated model to analyze such situations.~\cite{Galam1997,Michard2005,Phan2006,Sethna2001,Sethna2005,Phan2004} There are also other problems in which the opinions of the agents are represented as multidimensional continuous vectors and are influenced by the opinions of other agents.~\cite{Lorenz2007}

In this paper, we investigate the effect of social interactions on the consumption behaviors of people in a market. In a cultural market, there are many items competing with each other and the decision to consume an item or not is not independent of the consumption decisions of other items. The consumers have to pick one or more items from a wider pool of opportunities. 

An empirical study carried out by Salganik et al. (2006) provides experimental evidence that social influence has an effect on the consumption decisions of people.~\cite{Salganik2006} In the experiment, the subjects are faced with a web based application in which they can listen to and rate as many songs as they like among 48 songs of previously unknown bands. After they listen to and rate a song, they are offered the opportunity to download the song. The study reports the results of two different experimental conditions. The first one is called the independent condition in which the subjects only see the names of the songs without any other information and make their decisions independently from the other subjects. The second condition is called the social influence condition and in this case, the number of people who have downloaded each song so far is also given to the subjects. The number of downloads of a song is called the \textit{success} of the song. Any significant difference in the success of the same song between the two cases can be attributed to the availability of social information since there is no other experimental difference between the two settings.

The key finding of the study is that the availability of the social information significantly affects the way people behave. In the social influence condition, variation of the success outcomes of the songs are much higher than it is in the independent condition. This suggests that in the social influence case, some songs are downloaded many more times than they are in the independent case. Another measure they report is called the \textit{unpredictability} of a song and is found by calculating the average difference between the success values of a song over different realizations of the same condition (i.e. the experimental condition is repeated several times with different subjects). If a song tends to get the same outcomes over different realizations then its unpredictability value is low otherwise it is high. As a result of the experiment, the social influence condition leads to higher unpredictability values for the songs.

Borghesi and Bouchaud (2007) propose a generalization of the RFIM such that it allows to study multiple choices made by the agents simultaneously.~\cite{Borghesi2006} They show that it is possible to estimate some parameters of the model from the empirical data of Salganik et al.~\cite{Salganik2006}. Furthermore, they report that the behavior of the model changes qualitatively for low and high values of social pressure which is in alignment with the empirical findings.

Another line of research is carried out under \textit{social percolation} models.~\cite{Weisbuch2003} In this approach, consumers and producers are modeled as adaptive agents and the hit or failure of products are studied as a result of the information contagion between the consumers.
\label{cha:theory.introduction}


\section{Methodology}

\subsection{The Cultural Market Model}

\label{sec:cultural market}
The kind of market we are interested in is one where the items are easily reproducible so that their supply is practically unlimited and the consumers have to pick a subset of the available items to consume. An example of such a market is the cultural market where the items are cultural products such as books, movies, or music albums. Two major components of a cultural market are the set of consumers and the set of cultural items. Like Weisbuch and Solomon (2003), we use the movie market as an example in which the consumers are the customers of the movie theaters and items are the movies on the market.~\cite{Weisbuch2003}

An important assumption about a cultural market is that one of the factors effective in the consumption decision is the social pressure exerted on the individual. Theoretically, the effect may be both negative or positive but we focus on the case where it is positive (i.e. if an item is consumed more by the others, then it will have a greater chance to be consumed by the individual). The literature of psychology and economics have a large set of findings and different reasons for such effects.~\cite{Farrell1985} 

Formally speaking, the model consists of $N$ agents and $M$ items. We index the agents by Roman labels $i$ and $j$ (both in $[1,N]$) and the items by Greek labels $\alpha$ and $\beta$ (both in $[1,M]$). 

The agents are assumed to be located at the vertices of an undirected graph which incorporates the social structure of the community. If there is a link between agent $i$ and agent $j$ then these two agents are said to be neighbors of each other.

Initially, none of the items are consumed. The model advances in discrete time steps and each time step the agents form their opinions about the items. The \textit{opinion} of an agent $i$ about an item $\alpha$ is represented by $o_{i\alpha} \in \mathbb{R}$. At each time step, each agent picks the item with the highest opinion value and consumes it. An agent can consume an item only once so the selection is carried over non-consumed items. For each agent and item pair, $o_{i\alpha}$ is calculated as a weighted average of personal preferences and social pressure as $o_{i\alpha} = \gamma s_{i\alpha}+(1-\gamma)l_{i\alpha}$.

Social component $s_{i\alpha}$ represents the social pressure exerted on the agent $i$ to consume the item $\alpha$. In our model, we use a simple function which computes, for each agent $i$, the ratio of the number of neighbors who have consumed the item $\alpha$ to the total number of neighbors. Note that $s_{i\alpha} \in [0,1]$. Such a definition of social pressure is in accordance with our previous discussions since an agent will be more inclined towards consuming an item if a majority of its neighbors have already consumed it. Note that, in reality the social pressure is a function of time ($s_{i\alpha}=s_{i\alpha}(t)$) because as the agents consume items its value changes. Our implementation of the model is a synchronized one in the sense that at the start of a time step the social pressure is calculated according to current state and updated only when the time step is completed.

Personal preferences $l_{i\alpha} \in \mathbb{R}$ are predetermined and time independent values representing the likings of agents for the items. The value $l_{i\alpha}$ corresponds to the idiosyncratic personal taste of agent $i$ for the item $\alpha$. Higher values indicate higher preferences for the item.

The value $\gamma \in [0,1]$ is the \textit{social pressure parameter} which determines the strength of the social pressure on the decision process. The case $\gamma=0$ corresponds to the pure-individualistic community where no agent cares about what others are doing (hence basing their decisions solely on their idiosyncratic liking values), $\gamma=1$ corresponds to the pure-social environment where all decisions are based on the behaviors of others.

\subsection{Initial Configuration}

\label{sec:initial configuration}

\textit{Topology:} For the base case, the topology of the agents is set to a ring lattice with a coordination number of $k$ (i.e. agents are located in a ring topology and each agent is connected to its nearest $k$ neighbors). But the effect of employing other topologies is also investigated.

\textit{Personal preferences}: Personal preference data isolated from any social effect is very hard to obtain and no publicly available dataset was known to us at the time of this study. We represent the personal preference of the agents for the items as real numbers and assume that the idiosyncratic personal liking values $l_{i\alpha}$ come from a normal distribution with mean $\mu_\alpha$ and standard deviation $\sigma_{\alpha}$. This is an assumption that can be validated (or invalidated) in the future. Without any prior knowledge, assuming the liking values for the items come from a normal distribution specific to each item seems appropriate.

To simplify the matters more, we set the mean values of the distributions equal to each other (i.e. $\mu_\alpha = \mu$ where $\mu$ is a model parameter) due to our assumption that no item has any intrinsic superiority in terms of liking values over another item a priori.  Although this seems unrealistic, such an assumption can only weaken the inequalities that we observe as a result of the simulations and does not pose a problem to the conclusions made. Furthermore, it is trivial to show that the particular value of the parameter $\mu$ does not change anything in the dynamics of the model hence we can set $\mu=0$. Since only the ranking of the items are considered in the decision of the agents, an initial configuration with a non zero $\mu$ value can be safely turned into a zero-$\mu$ configuration by subtracting $\mu$ from every $l_{i\alpha}$.

The value $\sigma_{\alpha}$ is called the \textit{intra-item liking deviation} of item $\alpha$ and denotes the standard deviation in the liking values of agents for the same item. As a further simplification for the model, we assume all items have the same intra-item deviation and let $\sigma_{\alpha} =\sigma$ for all $\alpha$ where $\sigma$ is a model parameter. 

\subsection{Remarks}

We assume that consumption rates of the agents are constant and same for all. That means at each time step, each agent consumes exactly one item. Naturally, with these assumptions, all agents will consume all items after $M$ steps. In our model, a cultural item is available on the market until the end of the simulation. In real world however, not all items stay available in the market for an indefinite time period. Considering the movie market example, not all movies are shown in the theaters long enough to allow all consumers to see them. There is a continuous entrance and exit of movies. In order to keep our model simple, we ignore this fact but decide to terminate the simulations much before the point of total consumption  (i.e. after $T$ steps with $T<M$). In other words, our model is limited to a simplified version of the real world where $M$ items are put on the market at the same time and none of them leaves the market until the simulation ends. In addition, we assume that a single agent cannot consume all available items on the market but a small fraction of them.

We have seen that the liking values $l_{i\alpha}$ are chosen from a normal distribution with $\mu = 0$ and the same $\sigma$ for all $\alpha$. The fact that the expected average value of the liking values of an item over all agents is 0 does not imply that the sample mean of its liking values will be zero. Let $\itemquality = \sum_{i=1}^{N}l_{i\alpha} / N$ denote the sample mean of the liking values of item $\alpha$ over all agents. We call $\itemquality$ as the \textit{quality} of item $\alpha$. The Central Limit Theorem states that the distribution of the quality values will follow a normal distribution with mean 0 and standard deviation $\sigma/\sqrt{N}$ for large $N$.~\cite{Ross2002} In the limit $N\rightarrow\infty$ the deviation of the quality values approaches to 0. Although the model assumes no a priori superiority of an item over another, we should expect to see different item quality values for different realizations of the model for finite and small $N$.

\subsection{Quantities of Interest}
We define the \textit{market share}, $\marketshare$, of an item $\alpha$ as the ratio of consumers who have consumed that item so far. It is calculated as $\marketshare = \sum_{i=1}^{N}c_{i\alpha}/N$ where $c_{i\alpha}$ is 1 if agent $i$ has consumed the item $\alpha$ before the current time step and to 0 otherwise. Note that the market share of an item is defined such that the sum of the market shares of all items at a given time does not have to sum up to 1.

Item quality, $\itemquality$, is already defined as the average liking value of an item over all agents for an item $\alpha$.

\textit{Market Inequality}, $I$, represents the difference between the market shares of the cultural items. As a measure of the inequality, we calculate the Gini index of the market shares of all items. The market inequality is defined as follows.

\begin{equation}
I=\frac{(\sum_{\alpha=1}^M{\sum_{\beta=1}^M|\marketshare[\alpha] - \marketshare[\beta]|}) / M^2}{(2\sum_{\alpha=1}^M \marketshare)/M}
\end{equation}
which can be interpreted as half the relative expected difference between the market shares of two randomly chosen items.~\cite{Salganik2006} A perfectly equal market where all items have the same market shares will have an inequality value of 0; that is the Gini index of a perfectly equal market.

The market inequality value is a symmetrical measure with respect to the quality values of the items. It does not differentiate whether it is the low quality items that receive unfairly high market shares or the high quality items. In order to answer the question of whether there is a general trend favoring the high quality items in terms of market shares or vice versa, we employ a simple measure that is used in statistics which we call the quartile difference. First, we divide the items into four quartiles according to their quality values. The items with the top 25\% quality values are placed in the upper quartile $\mathcal{U}$ and the items with the bottom 25\% quality values are placed in the lower quartile $\mathcal{L}$. \textit{Quartile difference}, Q, is simply the difference between average market shares of the two quartiles. Since the number of items in the upper and lower quartiles are one forth of $N$ ($|\mathcal{U}|=|\mathcal{L}|=N/4$), quartile difference is defined as

\begin{equation}
Q = \sum_{\alpha\in \mathcal{U}}\frac{\marketshare}{N/4} - \sum_{\alpha\in \mathcal{L}}\frac{\marketshare}{N/4}
\end{equation}

The quartile difference is a signed value so the minimal value it can take is -1 and this occurs in the extreme case in which all items in the lower quartile have market shares of 1 and all the items in the upper quartile have market shares of 0. Similarly, the maximal value it can assume can be computed as +1.

\section{Results and Discussions}

All results reported in this section are based on values obtained by averaging over 100 independent runs of the simulation with the same parameters.

\begin{figure}[t]

\centering
\subfigure[] 
{
    \label{fig:quality-vs-share-a}
    \includegraphics[scale=0.26]{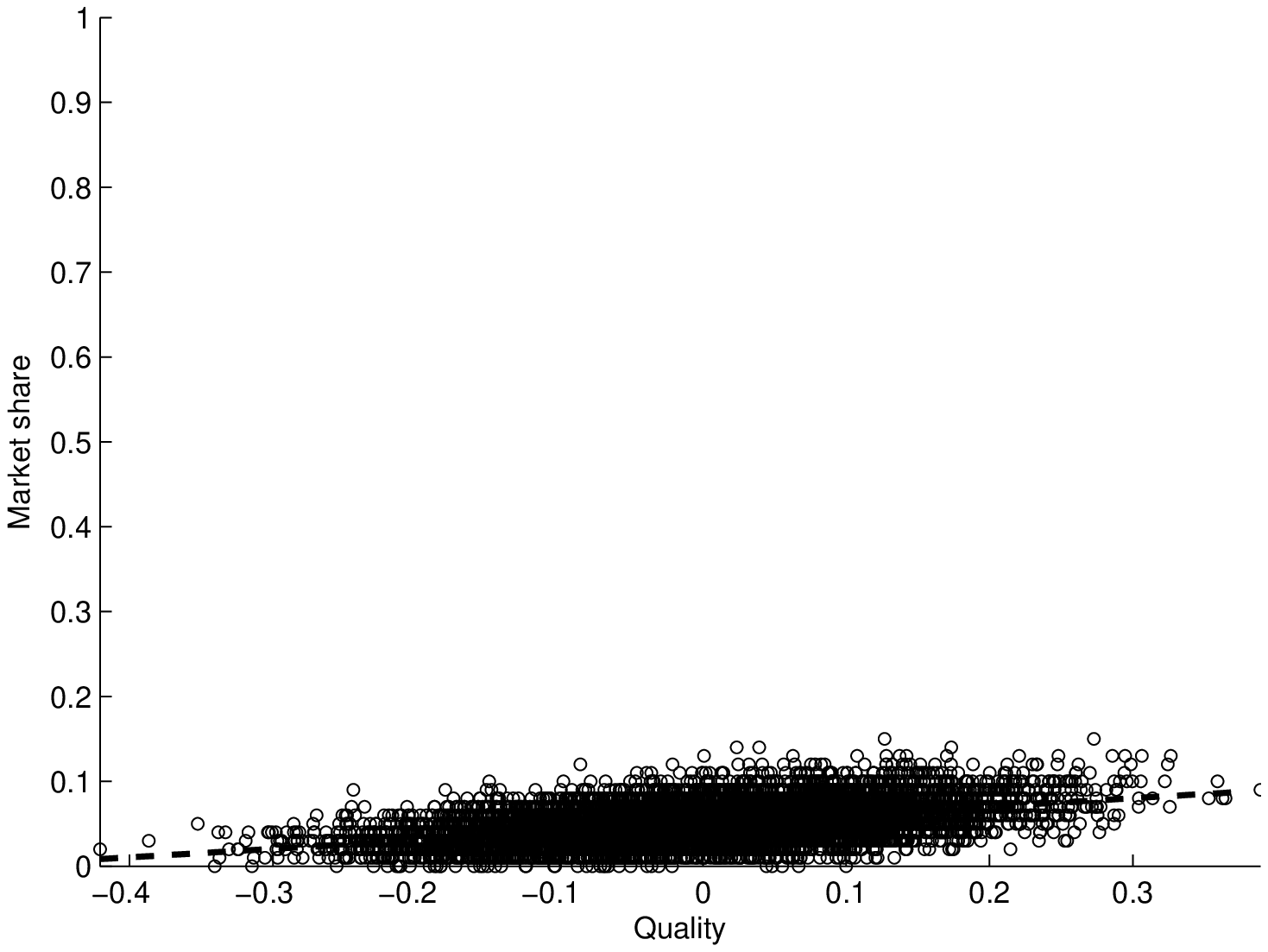}
}
\subfigure[] 
{
    \label{fig:quality-vs-share-b}
    \includegraphics[scale=0.26]{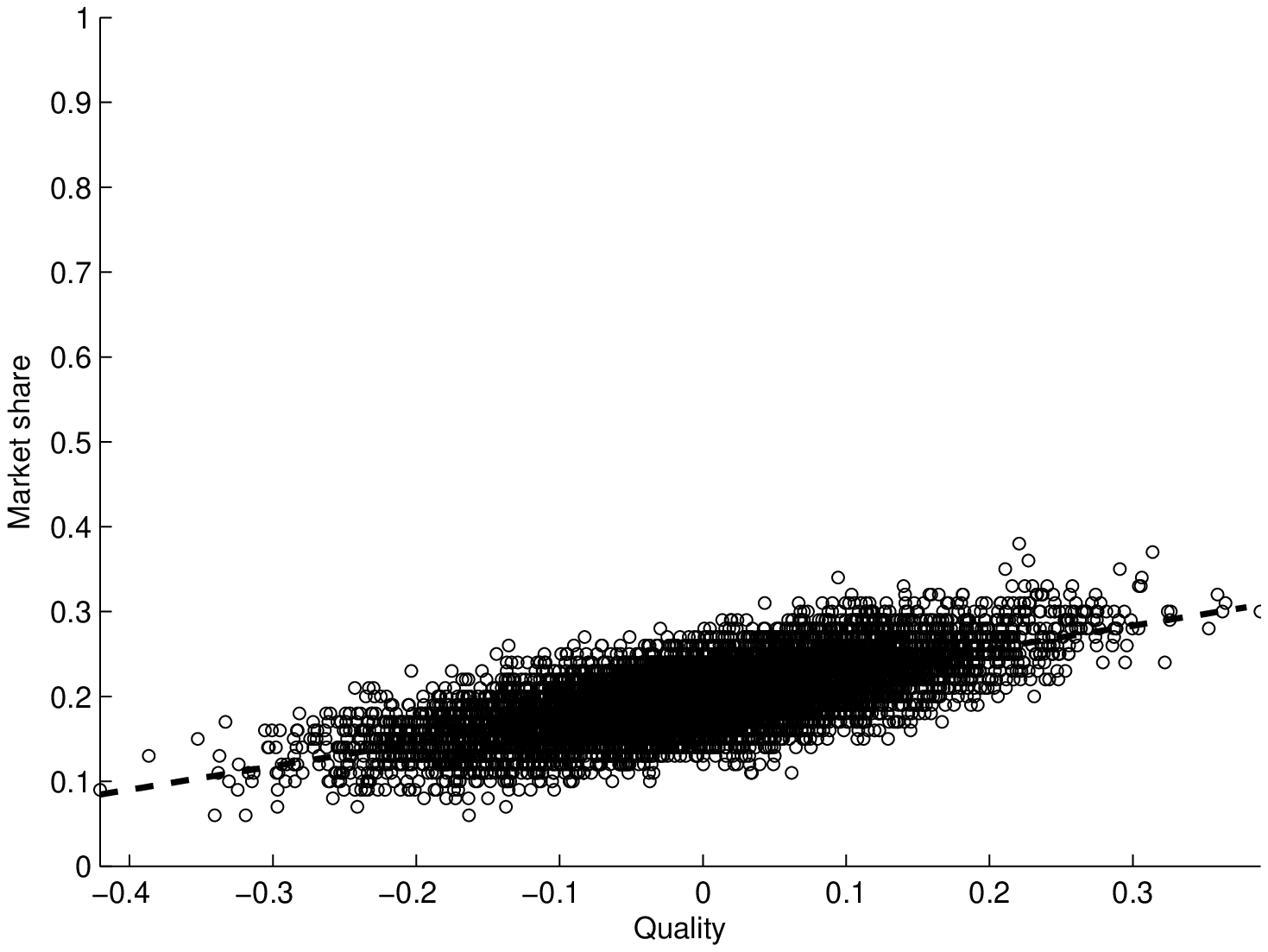}
}
\subfigure[] 
{
    \label{fig:quality-vs-share-c}
    \includegraphics[scale=0.26]{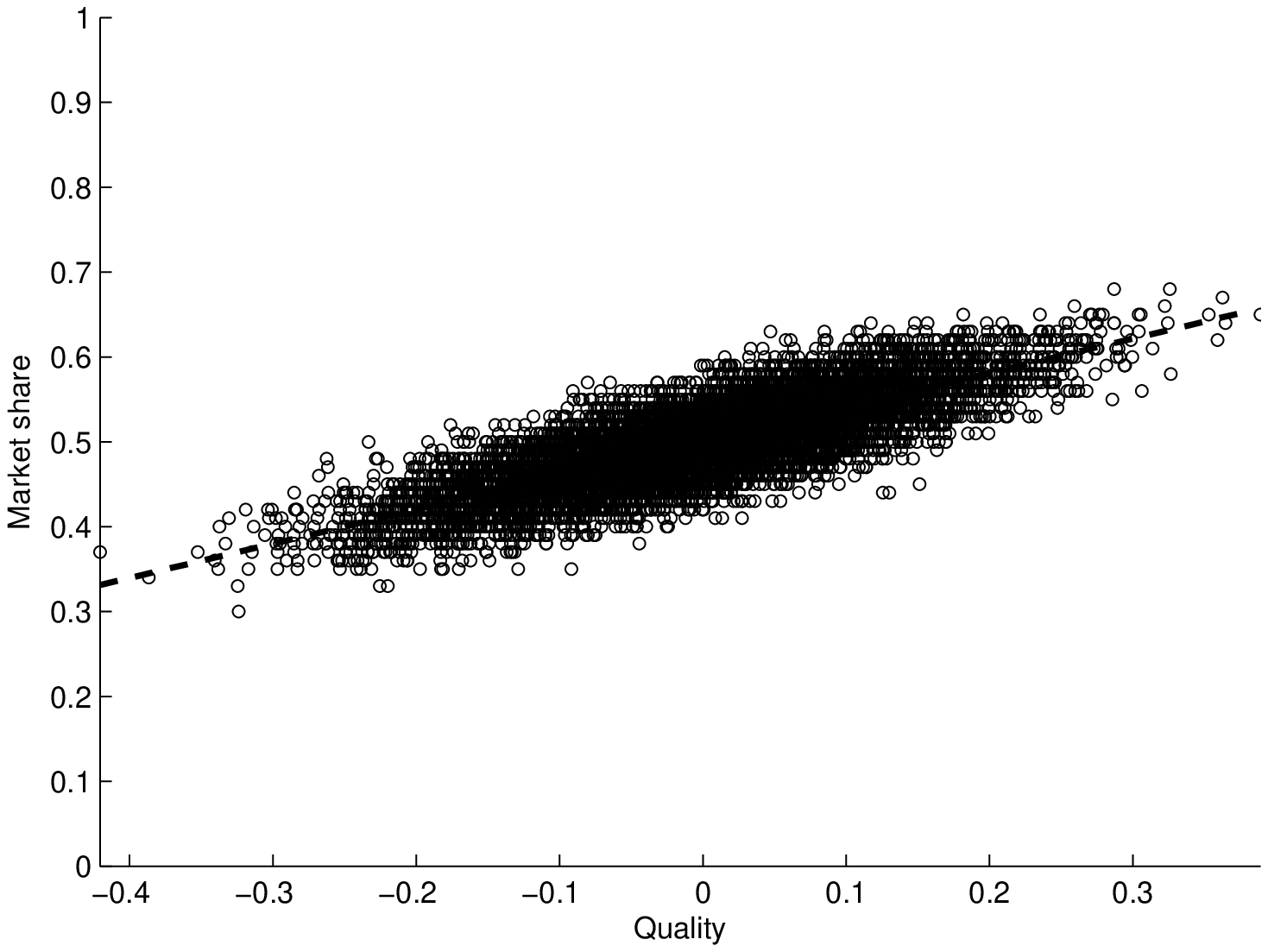}
}
\caption{Market shares versus quality values for no social pressure. The simulation length $T$ is (a) 5, (b) 20, and (c) 50.}
\label{fig:quality-vs-share} 
\end{figure}

It is natural to ask if the quality of an item determines its market share at the end or not. A reasonable expectation about a cultural market is that items with high quality values should get higher market shares on the average. As a starter, we let $\gamma=0$ and  $\sigma =1$ which corresponds to the pure-individualistic community where all agents base their consumption decisions solely on their personal tastes. We keep the number of agents and the number of items fixed to 100 in this set of experiments (i.e. $N=100$ and $M=100$). The sensitivity of the results on these parameters will be investigated separately. Since there is no social interaction, the underlying topology has no effect on the results because its possible effects are ruled out with the zero social pressure parameter (i.e. $\gamma=0$). We let the model run for 5, 20 and 50 steps (i.e. $T \in \{5,20,50\}$). In Fig.~\ref{fig:quality-vs-share}, we see the scatter plot of quality versus market shares of the items for three cases.

Let us first consider the case with $T=5$ given in Fig.~\ref{fig:quality-vs-share-a}. The relation between the quality and market share has a linear form. A linear relation also holds for the other two cases with $T=20$ and $T=50$ given in Fig.~\ref{fig:quality-vs-share-b} and \ref{fig:quality-vs-share-c} correspondingly.

Prolonging the simulation increases both total consumption and market shares. This observation is trivial given the model definition and can be verified by noting the data points are shifted up for higher $T$ values. Another expected result is that high quality items receive higher market shares and this is reflected in the positive slope values of the data points for all figures. A last observation is that the rate of market share increase differs with respect to the quality values of the items. The slopes, $m$, of the best fitting lines are significantly different for different values of $T$: $m_{T=5}=0.10$, $m_{T=20}=0.28$, and $m_{T=50}=0.40$. This means that high quality items increase their market shares at an higher rate than the low quality items.

As we already noted before, the simulation length (i.e. number of steps that the simulation will run for) is an important parameter and should be set carefully with keeping in mind the actual market to be modeled. It is obvious that values of $T$ very close to the number of items (i.e. 100 at the moment) are highly unrealistic. We set $T=20$ during the next set of experiments for simplicity of analyses. Such a setting leads to a market where each agent consumes 20 percent of all available items. Even though a choice of 20 seems arbitrary, further simulations revealed that as long as the simulation length is appreciably smaller than the number of items, the market shares and the quartile difference values do not change abruptly for differing values of $T$.~\cite{Herdagdelen2007} Unfortunately, the plots of these simulations are omitted in this paper due to space restrictions.

\begin{figure}[t]
\centering
\subfigure[] 
{
    \label{fig:share-share-low}
    \includegraphics[scale=0.40]{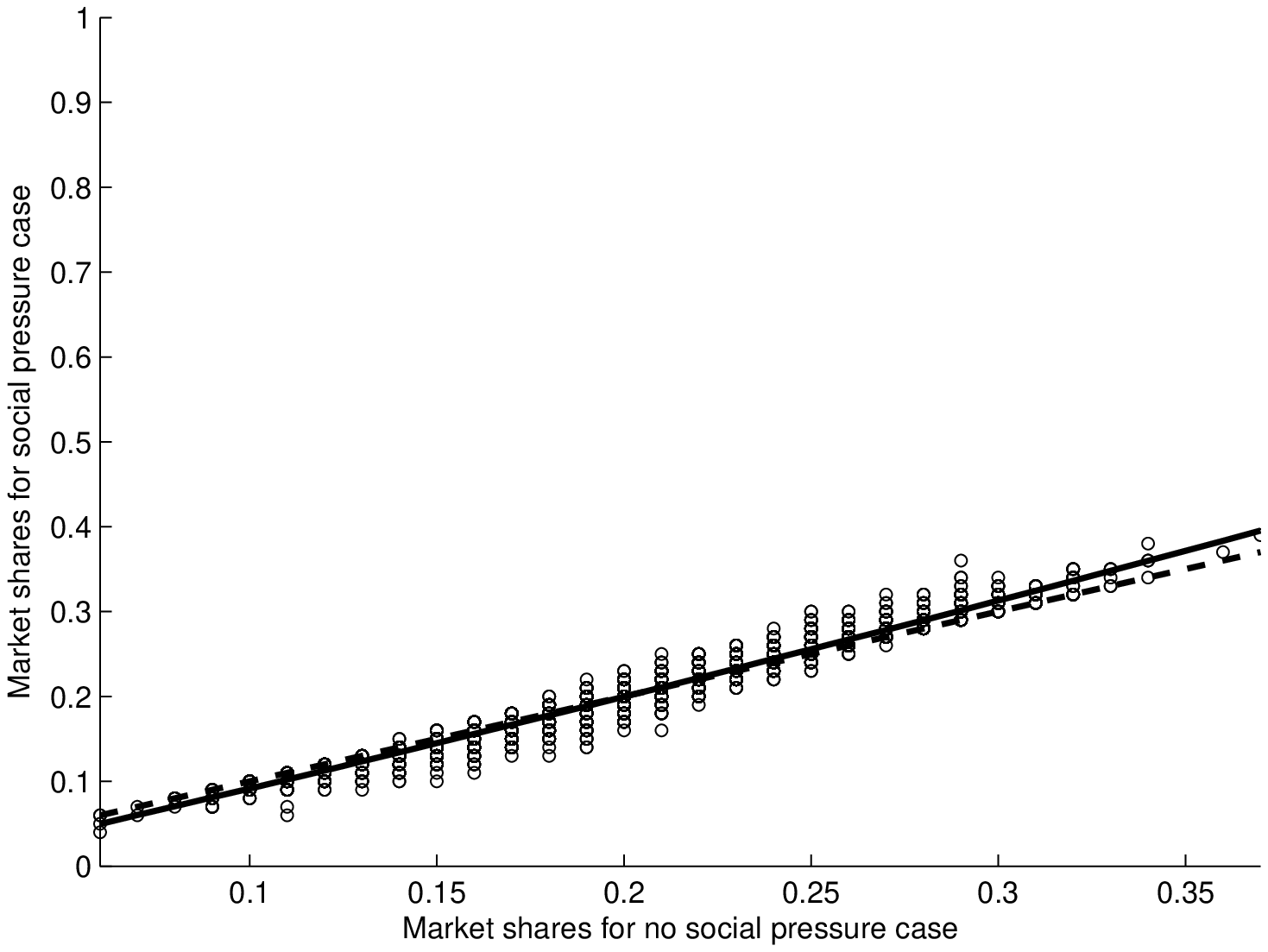}
}
\subfigure[] 
{
    \label{fig:share-share-high}
    \includegraphics[scale=0.40]{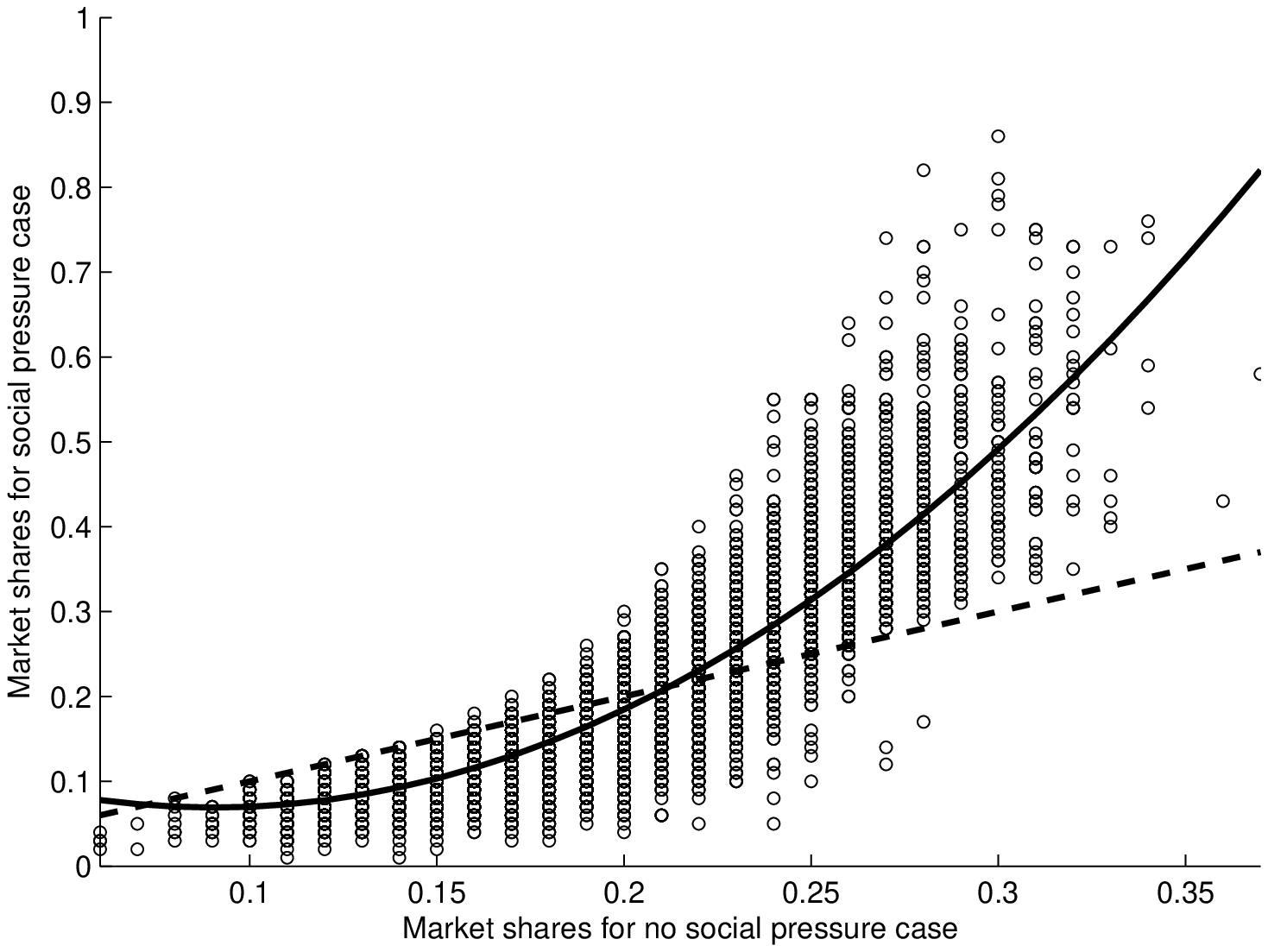}
}
\caption{The effect of introducing social pressure on market shares. (a) Low social pressure versus no social pressure.  (b) High social pressure versus no social pressure. In both figures, the dashed lines are the $y=x$ lines.}
\label{fig:share-share} 
\end{figure}

An interesting extension will be introducing social pressure (i.e. $\gamma>0$). To keep things simple for the moment, we create two parameter settings: One with $\gamma=0.3$ corresponding to low social pressure and one with $\gamma=0.7$ corresponding to high social pressure. With the introduction of social pressure, the topology of the neighborhood relations becomes important. For the moment, we set the topology to a fully connected graph which corresponds to a ring lattice with a coordination number $k=N-1$. An informative way to look at the effect of social pressure is to compare the market shares obtained for low and high social pressure to the ones obtained in the community with no social pressure. Using the same personal preference values allows us to compare the market shares of the same item under different social pressure environments. In order to compare the different social pressure settings, we present two figures (Fig.~\ref{fig:share-share-low} and Fig.~\ref{fig:share-share-high}) which contain the scatter plots of the market shares of the items obtained for low social pressure versus for no social pressure and high social pressure versus no social pressure correspondingly.

It is also possible to look at the scatter plots of quality and market shares obtained in these social settings. Since we have seen that quality and market shares are linearly related for no social pressure, this comparison also helps us to see the relation between quality and the market shares obtained in social pressure settings. These plots are omitted in this paper but they are in perfect accordance with the market share comparison graphs.

We obtain similar results for low social pressure and no social pressure. The data points are scattered around the $y=x$ line in Fig.~\ref{fig:share-share-low} indicating that the effect of setting $\gamma=0.3$ has no or limited effect on the final market shares of the items. For low social pressure, the consumption decisions remain fairly intact compared to no social pressure.

For high social pressure, the relation between the quality and market share of an item is no longer linear as Fig.~\ref{fig:share-share-high} suggests. The items which obtain high market shares for no social pressure obtain even higher market shares for high social pressure and the items which obtain low market shares for no social pressure obtain even lower market shares for high social pressure.

These results are compatible with the experimental data obtained in the study of Salganik et al. (2006) (see Fig 3 in Ref.~\refcite{Salganik2006}). In this experimental study, the introduction of social information causes a non linearity between the market shares obtained with the presence of social pressure or no social pressure. Our model is able to generate this phenomenon.

What is the response of our model to varying degrees of social pressure? How can we extend our findings for communities with varying degrees of heterogeneity? The model parameter $\gamma$ regulates the extent of social pressure and intra-item liking deviance $\sigma$ regulates the heterogeneity of the agents in the community. Until now, we set $\sigma =1$ but this choice is arbitrary. Smaller choices for $\sigma$ will lead to more homogeneous communities in the sense that the liking values of the agents for the same items will be closer to each other. Higher choices on the other hand, will correspond to more heterogeneous communities because the liking values of the agents for the same items will deviate more.

\begin{figure}[t]
\centering
\subfigure[]{
\label{fig:sigma-gamma-series-I}
\includegraphics[height=4.5cm]{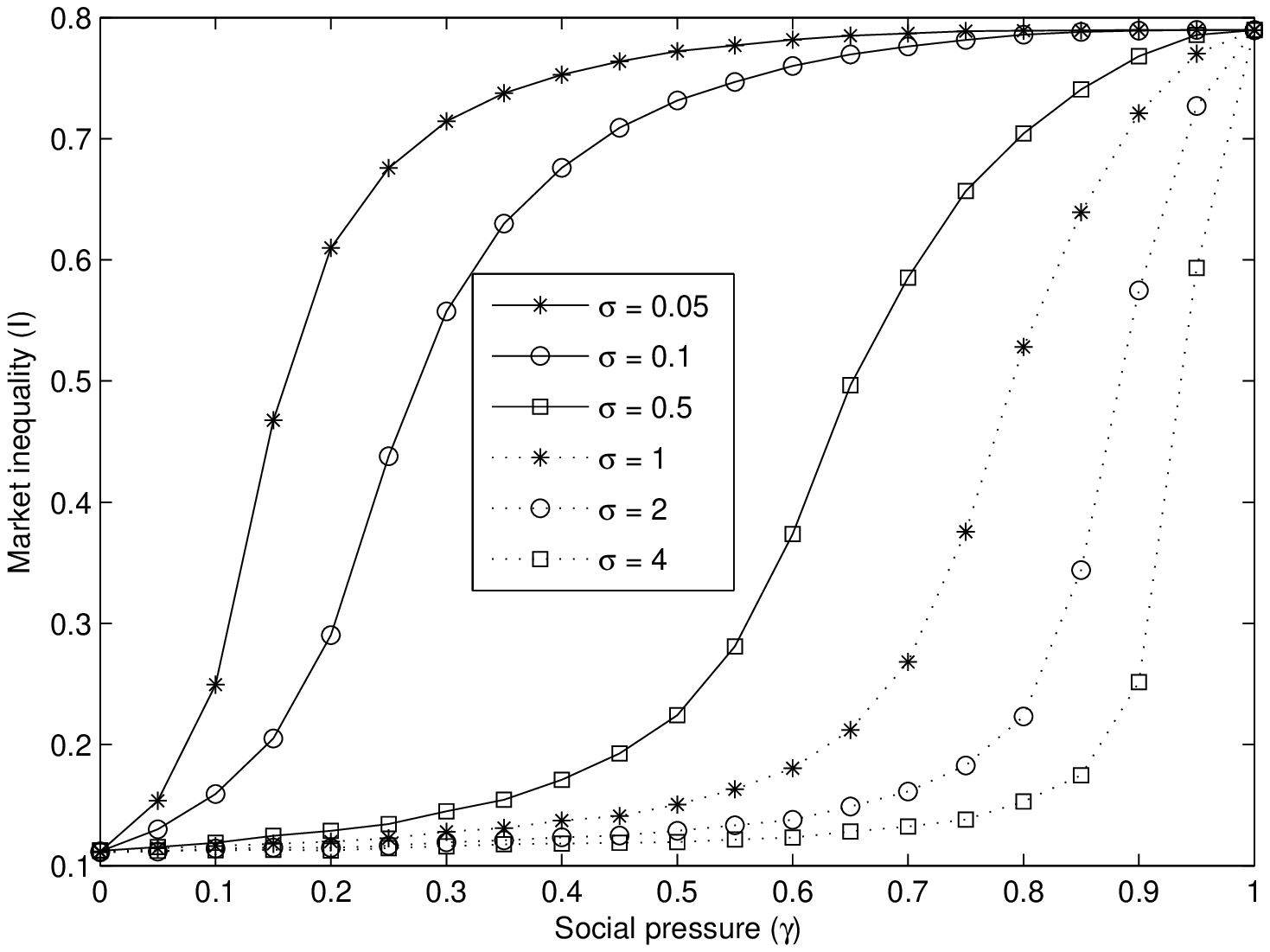}
}
\subfigure[]{
\label{fig:sigma-gamma-series-Q}
\includegraphics[height=4.5cm]{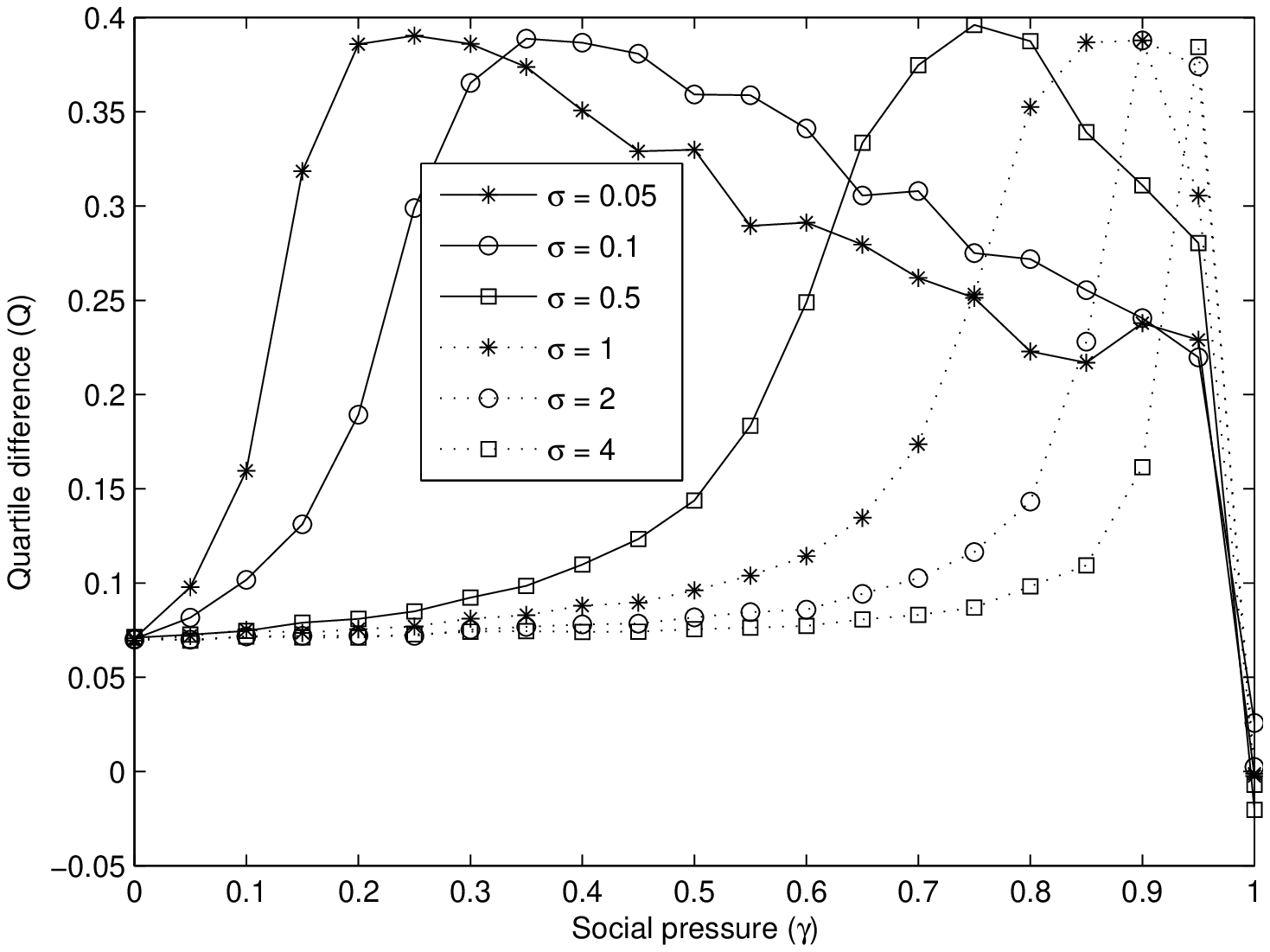}
}

\caption{Effect of intra-item liking deviation $\sigma$ (line series) and social pressure $\gamma$ on (a) the inequality of the market shares $I$, (b) quartile difference Q.}

\label{fig:sigma-gamma-series} 

\end{figure}

In order to come up with answers to the two questions we ask, we calculated the market inequality $I$ and the quartile difference $Q$ of the markets at the end of 20 steps for different pairs of $\gamma$ and $\sigma$ values. Fig.~\ref{fig:sigma-gamma-series} visualizes the differing values of $I$ and $Q$ correspondingly.

In Fig.~\ref{fig:sigma-gamma-series-I}, we see the effect of increasing social pressure $\gamma$ on the market inequality for different values of $\sigma$. As we have already seen previously, higher $\gamma$ values lead to higher market inequality values but the characteristic of the effect depends on $\sigma$. We see that the marker inequality values show a sharp increase from low values (e.g. $I=0.2$) to high values (e.g. $I=0.8$) at critical values of $\gamma$. However the specific value of the critical $\gamma$ value depends on $\sigma$.

The market inequality is a measure of how varied the final market shares are. It does not tell us anything about the dependence of market shares on quality. We have already seen that higher social pressure results in increased market shares for the high quality items. Figure \ref{fig:sigma-gamma-series-Q} visualizes the effect of increasing social pressure on the quartile difference $Q$ for different values of intra-item liking deviation. Similar to the market inequality value, the quartile difference also shows a sharp increase at a critical value of $\gamma$ and the specific value of $\gamma$ depends on $\sigma$. However, at higher levels of social pressure the quartile difference starts to decrease. The turning point for the quartile difference (i.e. the $\gamma$ value that it starts to decrease) is again dependent on $\sigma$. This is an interesting observation because after the turning point the inequality value continues to increase while the quartile difference starts to decrease. Considering the extreme case where $\gamma=1$ may help to understand this phenomenon. When $\gamma=1$, the agents will not consider the liking values in their decision process but only the number of neighbors who have consumed the items. Thus, the quality values of the items will be irrelevant in affecting their market shares leading to a quartile difference value of zero.

Interpreting the results for inequality and quartile difference values together we conclude that there exists a critical region of $\gamma$ and for values less then the critical region the inequality and quartile difference values are in accordance with each other. High quality items gain higher market shares as we increase the social pressure and this fact is reflected in increased inequality and quartile difference values. But once the critical turning point in social pressure is passed, the relation between the quality and market share starts to weaken, but some items still continue to arbitrarily high market shares at the cost of others. Hence the inequality value continues to increase. The positive correlation between the quality and market shares disappears suddenly as reflected by the decrease in quartile difference value.

In order to assess whether the reported findings are robust with respect to the model parameters, we repeated the simulations with different parameter setting. Due to space limitations, we are unable to give a detailed report of those simulations but the results are given in an unpublished masters thesis.~\cite{Herdagdelen2007}

We repeated the simulations for $N\in\{100, 1000, 5000, 10000\}$ and $M\in\{100, 1000, 5000\}$ and the results were qualitatively the same with the results we report in this study. Also the effect of the number of neighbors $k$ and the topology is investigated. Even for very low values of $k$ and random topology with directed links, we did not observe qualitative differences compared to ring lattice.

\section{Conclusions and Future Work}

Empirical findings suggested that the introduction of social interaction has a profound effect on the relation between the quality and market share of an item. For low values of social pressure (i.e. $\gamma=0.3$), we observed that the linear relation between the quality and market shares remains almost intact. As we increase the social pressure parameter ($\gamma=0.7$), the linearity was disrupted and we observed nonlinear relation in favor of high quality items.

An interesting finding is that the social pressure parameter does not regulate the collective nature of the model but interacts with $\sigma$ which represents the deviation among the liking values of the agents for the same items. We found out that a given market inequality value $I$ can be reproduced by a set of ($\gamma$, $\sigma$) pairs and the interaction between the two parameters remain qualitatively intact for different number of agents $N$ and average number of neighbors $k$.

To our surprise, while the inequality in the market shares increases with increasing social pressure, the quartile difference value first increases and than decreases with a peak at a critical value of social pressure dependent on $\sigma$.  As we increase social pressure, the high quality items gain more and more market shares at the cost of low quality items and this is reflected in the increased inequality and quartile difference values. But after the critical social pressure value is passed, the positive correlation between the quality and market share deteriorates. Low quality items starts to gain higher market shares and quartile difference shows a decrease. Note that we cannot observe this trend just by looking at the market inequality which continues to increase as we increase the social pressure. These findings suggest that for higher values of social pressure the dynamics of the model gets history dependent and the resulting market share of an item is determined by the initial conditions of the system rather than the quality of the item.

We carried out extended simulations to see if our results depend on specific values of the parameters or robust to different values of the parameters. We concluded that the qualitative nature of the simulations are robust with respect to different number of agents (i.e. $N \in \{100, 500, 1000, 5000\}$) and varying degrees of network connectivity (i.e. for different values of $k<N$). The topology on the other hand seems to be affecting the outcome of the cultural market. The sharp increases in the inequality and quartile difference were more resilient to varying number of agents in random topology than the ring topology.

We studied the ring topology as a base model and introduced the heterogeneity by using random topology. How the model will behave if another type of network such as scale free or small world is introduced is definitely an interesting and non-trivial question and needs to be addressed in the future studies. Also the effect of introducing directed links and asymmetrical neighboring relations can be investigated in the future.

Another important point is to study the accordance between the real world data and the model output. Unfortunately, during the writing of this paper, no relevant dataset was publicly available. With the limited information available from a previous experiment (Ref.~\refcite{Salganik2006}) we were able to show that our model can produce compatible results and it can provide an explanation for the observed differences in the market share, item quality relations and market inequality values depending on the availability of social information. Additionally, we believe that study of such computational models is meaningful given the costs of conducting a real world experiment is many times higher than building and analyzing a computational model. The results and the dynamics of the cultural market model can be of use in determining the settings of a real world experiment in the future.

\section{Acknowledgments}

This work was partially supported by Bogazici University Research Fund under the grant number 07A105 and is partially based on the work performed in the framework of the FP6 project SEE-GRID-2, which is funded by the European Community (under contract number INFSO-RI-031775).


\end{document}